\documentclass[doublecol]{epl2}
\usepackage{graphicx}
\usepackage{dcolumn}
\usepackage{bm}

\title{Simulation of temperature profile for the electron- and the lattice-systems in laterally structured layered conductors}
\shorttitle{Simulation of temperature profile for the electron- and the lattice-systems in laterally structured layered conductors} 

\author{L. Yang\inst{1,2} \and R.J. Qian\inst{1,2} \and Z.H. An\inst{1,3,(a)} \and S. Komiyama\inst{2,4,5} \and W. Lu\inst{2,6}}
\shortauthor{L. Yang \etal}

\institute{
  \inst{1} Key Laboratory of Surface Physics, Institute of Nanoelectronic Devices and Quantum Computing, Key Laboratory
  of Micro and Nano Photonic Structures (Ministry of Education), Department of Physics, Fudan University - Shanghai 200433, China\\
  \inst{2} National Laboratory for Infrared Physics, Shanghai Institute of Technical Physics, The Chinese Academy of Sciences - Shanghai 200083, China\\
  \inst{3} Collaborative Innovation Center of Advanced Microstructures - Nanjing 210093, China\\
  \inst{4} Terahertz Technology Research Center, NICT, Nukui- Kitamachi 4-2-1, Koganei, Tokyo, Japan\\
  \inst{5} Department of Basic Science, The University of Tokyo - Komaba3-8-1, Meguro-ku, Tokyo, 153-8902, Japan\\
  \inst{6} School of Physical Science and Technology, ShanghaiTech University - Shanghai 201210, China\\
}

\abstract{
 Electrons in operating microelectronic semiconductor devices are accelerated by locally varying strong electric field to acquire effective electron temperatures nonuniformly distributing in nanoscales and largely exceeding the temperature of host crystal lattice. The thermal dynamics of electrons and the lattice are hence nontrivial and its understanding at nanoscales is decisively important for gaining higher device performance. Here, we propose and demonstrate that in layered conductors nonequilibrium nature between the electrons and the lattice can be explicitly pursued by simulating the conducting layer by separating it into two physical sheets representing, respectively, the electron- and the lattice-subsystems. We take, as an example of simulating GaAs devices, a 35nm thick $1\mu m$ wide U-shaped conducting channel with 15nm radius of curvature at the inner corner of the U-shaped bend, and find a remarkable hot spot to develop due to hot electron generation at the inner corner. The hot spot in terms of the electron temperature achieves a significantly higher temperature and is of far sharper spatial distribution when compared to the hot spot in terms of the lattice temperature. Similar simulation calculation made on a metal (NiCr) narrow lead of the similar geometry shows that a hot spot shows up as well at the inner corner, but its strength and the spatial profiles are largely different from those in semiconductor devices; viz., the amplitude and the profile of the electron system are similar to those of the lattice system, indicating quasi-equilibrium between the two subsystems. The remarkable difference between the semiconductor and the metal is interpreted to be due to the large difference in the electron specific heat, rather than the difference in the electron phonon interaction. This work will provide useful hints to deeper understanding of the nonequilibrium properties of electrical conductors, through a simple and convenient method for modeling nonequilibrium layered conductors.}

\begin{document}

\maketitle

\section{Introduction}
Electro-thermal behavior is a key ingredient for understanding charge carrier transport phenomena in semiconductor devices including two-dimensional (2D) materials, hetero Junctions, and strong correlated systems \cite{Aninkevicius.V,Matulionis.A,Levi.A,Atar.F,Pumarol.M,Lan.Y,Guo.Z,Lee.S,Zhang.P,Li.Z,Lee.H,Lee.Y,Shao.W,Konstantinova.T,Wu.K}. In small devices on nanoscales, hot electron generation and the resulting characteristic interaction with the host crystal lattice (or phonons) complicates the electro-thermal analysis and limits the device performance \cite{Tisdale.W,Ashalley.E}. Whereas knowing the detailed local profile of the electron effective temperature, $T_{e}$, separately from that of the lattice temperature, $T_{L}$, in the presence of current is prerequisite for understanding the transport characteristics on nanoscales \cite{Dubi.Y,Hartmann.M,Mahajan.R,Brongersma.M,Peng.S},$T_{e}$ has been experimentally hardly accessible \cite{Reparaz.J,Xu.Y,Zhang.X,Nonnenmacher.M,Harzheim.A,Yalon.E,Cui.L,Shi.L,Mecklenburg.M} until quite recently \cite{Weng.Q,Komiyama.S}. It follows that the study of electro-thermal properties has so far been restricted only to the simulation methods such as those of Monte Carlo (MC) simulation based on the Boltzmann transport equations, hydrodynamic equations or molecular dynamics \cite{SINHA.S,Vasileska.D,Fan.A,Garcia.S,Wang.Y}. Unfortunately, however, MC simulations comprise involved calculation procedures, which are not necessarily convenient to gain intuitive understanding of the electro-thermal transport phenomena of given devices. On the other hand, the nonequilibrium condition cannot be incorporated in commercially available semiconductor device simulators.

Here, we propose a simplified electro-thermal model for layered conductors on the basis of the assumption that the electron- and the lattice-subsystems are, respectively, in quasi-equilibrium states characterized by the effective electron temperature $T_{e}$ and the lattice temperature $T_{L}$. The model is applied to a U-shaped layered conductor, where electric field is concentrated at the inner corner of the U-shaped bend. In a semicodncuctor device, simulating GaAs, remarkable hot electron distribution ($T_{e}\gg T_{L}$) is found to develop at the corner, forming a sharp hot spot with $T_{e}$ reaching $\sim2000K$. Differently, in metal devices, simulating NiCr, hot electron effects are found to be absent ($T_{e} \approx T_{L}$), whereas a hot spot profile is visible. These findings are consistent with recent experimental results reported on metals \cite{Weng.QC} and semiconductors \cite{Weng.Q}, indicating the validity of the present model for simulating the electro-thermal behavior of layered conductors in nonequilibrium conditions.

\section{Simulation model}
Figure 1 describes the model conductor considered in this study. A layered conductor with the electric conductivity $\sigma_{e}$ is deposited on an insulating substrate, which is anchored by the heat sink at 300K. The lateral shape of the conductor is arbitrary, so that the electric field $\mathbf{E}$, the current density $\mathbf{j}$, the electron temperature $T_{e}$ and the lattice temperature $T_{L}$ in the conductor are variables to be consistently derived as functions of the lateral position $\mathbf{r}$ for a given conductor with a given bias voltage. In the conductor electrons gain energy from $\mathbf{E}$ through $P=\mathbf{j}(\mathbf{r})\cdot \mathbf{E}(\mathbf{r})=\sigma_{e}E^{2}$ and the energy gained from the field is, in turn, released to the lattice via electron phonon interaction, characterized by the electron-phonon energy relaxation time $\tau_{e-ph}$. The excess energy (or heat) of electrons is transferred, as well, within the electron system through the electron thermal conduction $-\kappa_{e}\bigtriangledown T_{e}$ with $\kappa_{e}$ the electron thermal conductivity. The heat is transferred similarly within the lattice system through lattice thermal conduction $-\kappa_{L}\bigtriangledown T_{L}$ with $\kappa_{L}$ the lattice thermal conductivity. The heat is eventually transferred to the substrate $(T_{L}-T_{LS})/h_{I}$ with $h_{I}$ being the interface thermal resistance and $T_{LS}(\mathbf{r})$ the local lattice temperature of the substrate on its top surface, and finally absorbed by the heat sink. Heat is transferred as well through electrical leads connected to the conductor, as represented by the arrows marked with $\kappa_{e}$ and $\kappa_{L}$ in fig.1 (c), which is taken into account in the model through an appropriate boundary condition as mentioned below for fig.2 (a).

\begin{figure}
\includegraphics[width=1.0\linewidth]{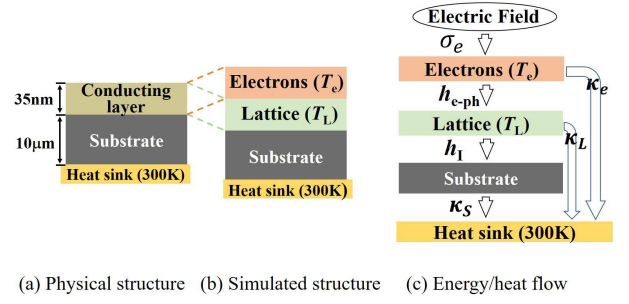}
\caption{Model for electro-thermal analysis. (a) Physical structure of the sample. (b) Simulated structure, in which the electron- and the lattice-systems of the conductor are separated to thermally contacted two layers. (c) Diagram of energy/heat flow.}
\label{fig.1}
\end{figure}

As schematically shown in figs.1 (b) and (c), our model represents the energy transfer from the electron system to the lattice system in the conductor by the interface heat transfer between the electron sublayer at $T_{e}$ to the lattice sublayer at $T_{L}$. The energy flux released from the electron system to the lattice system through the electron phonon interaction is given by

$$P_{e-ph}=(T_{e}-T_{L})C_{e}/\tau_{e-ph}\eqno{(1)}$$

with $C_{e}$ the electron specific heat per unit area, so that the effective interface thermal resistance $h_{e-ph}$ is
\
$$h_{e-ph}=\tau_{e-ph}/C_{e}.\eqno{(2)}$$
\
The specific heat is approximated by
\
$$C_{e}=C_{ec}=(3/2)N_{2D}k_{B},\eqno{(3)}$$
\
for a classical electron system ($k_{B}T_{e}\gg \varepsilon_{F}$) and by
\
$$C_{e}=C_{eF}=\{(3/2)k_{B}T_{e}/\varepsilon_{F}\}C_{ec}\eqno{(4)}$$
\
for an electron system with the Fermi energy $\varepsilon_{F}$ much higher than the thermal energy ($k_{B}T_{e}\ll \varepsilon_{F}$). Here, $k_{B}$ is the Boltzmann constant, and $N_{2D}$ is the 2D electron density.

\section{Simulated structure}

\begin{figure}
\includegraphics[width=1.0\linewidth]{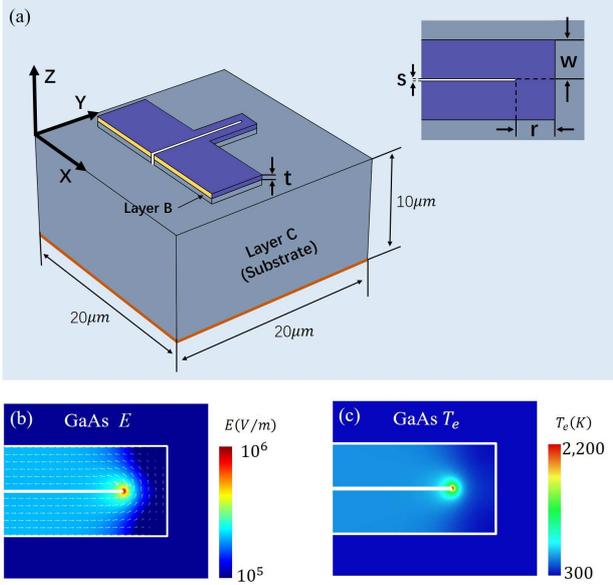}
\caption{(a) Simulated sample structure, where the electron- and the lattice-systems of a U-shaped conducting channel of a thickness $t=35nm$ are separately represented, respectively, by Layers A and B. The conducting channels with a width of $w=1\mu m$ $(r=1\mu m)$ and a length of $l=4\mu m$ extend from the two $5\times7\mu m^{2}$ contact pads. The radius of curvature of the inner corner of the U-shape is 15nm, making a gap of $S=30nm$ between the two channels. Bias voltage $V_{b}$ is defined as the voltage difference between the end faces of Layer A (marked by yellow). (b) and (c) Simulated electric field distribution and temperature distribution in the electron system (Layer A) of n-GaAs.}
\label{fig.2}
\end{figure}

As schematically illustrated in fig.2 (a), we consider a 35nm thick conductor layer shaped into a $1\mu m$-wide U-shaped channel with the radius of inner curvature of the U-shape is 15nm and the gap between the channels is $S=30nm$. For the simulation, the electron- and the lattice-systems of the conducting channel are separately represented by Layers A and B, where Layer B is placed on the $10\mu m$-thick substrate (Layer C). The boundary condition of temperature is given by assuming $T=300K$ on the bottom face Layer C and on the end faces of Layers A and B as marked by the orange lines in fig.2 (a). As to the bias condition, a constant voltage is assumed on each end face of the conducting channel (Layer A), and a bias voltage $V_{b}$ is assumed to give the voltage difference between the two end faces.

Two different conductors are considered. One is a doped n-GaAs channel and the other is a NiCr channel, similar to those studied, respectively, in Refs.32 and 39. Substrates are assumed to be lattice-matched GaAs/AlGaAs for n-GaAs sample \cite{Weng.Q} and single crystal Si covered with a thin $SiO_{2}$ layer for NiCr sample \cite{Weng.QC}. The electron density in n-GaAs and NiCr samples are, respectively, $N_{3D}=3.3\times10^{24}/m^{3}$ and $1.0\times10^{30}/m^{3}$; in terms of the sheet electron density, $N_{2D}=1.1\times10^{17}/m^{2}$ and $3.5\times10^{22}/m^{2}$. The specific heat is taken to be $C_{e}=2.3\times10^{-6}Ws/(Km^{2})$ and $1.1\times10^{-2}Ws/(Km^{2})$, respectively assuming Eqs. (3) and (4) for n-GaAs and NiCr samples. In n-GaAs sample, the interface thermal resistance ($h_{I}$) is negligibly small because the n-GaAs conducting layer is epitaxially grown on the lattice matched substrate. The electron-phonon energy relaxation time is assumed to be $\tau_{e-ph}=1ps$ and 3ps, respectively for n-GaAs \cite{Weng.Q} and NiCr \cite{Weng.QC}. Parameter values used are summarized in Table 1.

\begin{table*}[htbp]
\caption{Parameters used in the simulation.}
\label{tab.1}
\begin{center}
\begin{tabular}{|c|c|c|c|c|c|c|}
\hline
Quantity  & $\sigma_{e}$ & $\kappa_{e}$ & $\kappa_{L}$ & $h_{e-ph}$ & $h_{I}$ & $\kappa_{S}$\\\hline
Unit & $S/m$ & $W/(m\cdot K)$ & $W/(m\cdot K)$ & $Km^{2}/W$ & $Km^{2}/W$ & $W/(m\cdot K)$\\\hline
n-GaAs  & $8.8\times 10^{4}$ & 0.1 & 50 & $4.3\times10^{-7}$ & 0 & 50\\\hline
NiCr & $2.89\times 10^{5}$ & 15 & 1 & $2.7\times10^{-10}$ & $3\times10^{-8}$ & 150\\\hline
\end{tabular}
\end{center}
\end{table*}

\section{Results and discussions}
Joule heating caused by electric field and thermal conduction generated by temperature gradient or difference are consistently treated by using a commercial multiphysics software (COMSOL), where the bias voltage is taken to be $V_{b}=4.5V$ in all the calculations described below. It is a common feature of both n-GaAs and NiCr samples that the electric field is concentrated around the U-shaped inner corner as exemplified by the result for n-GaAs sample: Electric field is nearly uniform and $E=2\sim3kV/cm$ in a region away from the U-shaped corner, but rapidly increases to reach about $E=10kV/cm$ in the vicinity of the U-shaped inner corner. As a consequence of this $E$-field enhancement, remarkable nonuniform hot-electron distribution is found to be generated at the inner corner of the n-GaAs sample as shown in fig.2 (c). While the trend of the $E$-field enhancement is substantially the same in the NiCr sample, resulting temperature distribution in the electron- and the lattice-systems is largely different as described in detail below.

\begin{figure}
\includegraphics[width=1.0\linewidth]{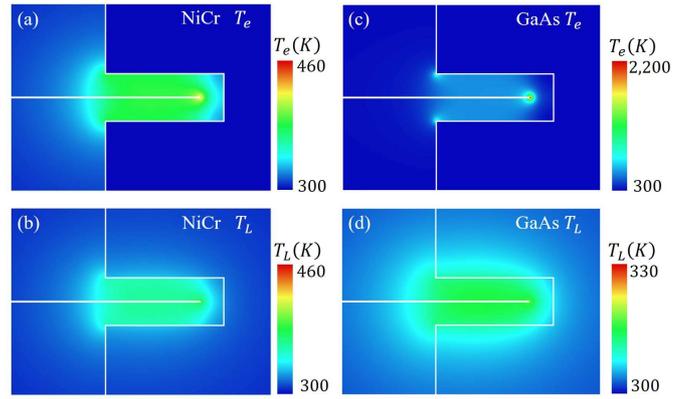}
\caption{(a) and (b) Temperature distributions of electrons and the lattice in the NiCr sample. (c) and (d) Temperature distributions of electrons and the lattice in the n-GaAs sample. White lines indicate the borders of conducting layers.}
\label{fig.3}
\end{figure}

Figures 3 (a), (b) and figures 4 (a)-(d) display the distributions of $T_{e}$ (Layer A) and $T_{L}$ (Layers B) for the NiCr sample. The profile of $T_{e}$  is similar to that of $T_{L}$, and both exhibit spatially varying heating in accord with the $E$-field enhancement peaked at the U-shaped inner corner. The highest temperature at the hot spot is about $150^{\circ}$C above the heat sink (300K). The amplitude of the temperature rise at the hot spot ($\bigtriangleup T_{e}\approx150^{\circ}$C) as well as the quasi-equilibrium feature between the electron- and the lattice-systems ($T_{e}\approx T_{L}$) substantially reproduce the experimental findings reported in Ref.39.

\begin{figure}
\includegraphics[width=1\linewidth]{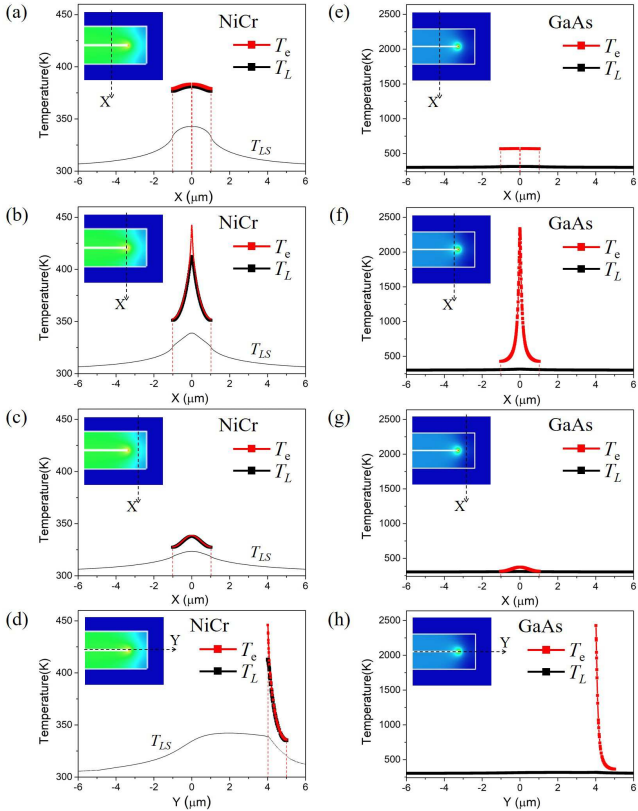}
\caption{One-dimensional plots of $T_{e}$ and $T_{L}$, along the black dashed arrows shown in the inset, for the NiCr sample (a)-(d) and for the n-GaAs sample (e)-(h).}
\label{fig.4}
\end{figure}

The feature of the hot spot formation is largely different in the n-GaAs sample as shown in figs.3 (c)(d) and figs.4 (e)-(h). The electron temperature $T_{e}$ is much higher than the lattice temperature $T_{L}$, indicating nonequilibrium hot electron generation, and it assumes a very sharp prominent peak reaching as high a value as $T_{e}\sim 2000K$. On the other hand, the highest value of $T_{L}$ ($<330K$) is at most only $\sim30^{\circ}$C above the temperature of the heat sink (300K). In addition, the hot-spot feature is practically missing as evident in figs.4 (f) and (h). The generation of remarkable hot electron distribution is consistent with the experimental finding reported on n-GaAs constriction devices \cite{Weng.Q}.

The large difference in the electro-thermal properties noted between the n-GaAs sample and the NiCr sample in this study is suggested to be generally inherent to semiconductors and metals. When energy flux $P$ is fed to the electron system in a steady state, the electrons are heated above the lattice temperature by
\
$$T_{e}-T_{L}=(\tau_{e-ph}/C_{e})P=h_{e-ph}P\eqno{(5)}$$
\
if temperature gradient is ignored. For a given $P$, the rise of $T_{e}$ is proportional to $\tau_{e-ph}$, and $1/C_{e}$. In general, $\tau_{e-ph}$ is not largely different between semiconductors and metals, but the heat capacity $C_{e}$ is by orders of magnitude smaller in semiconductors because the electron density is far lower. It follows that the electron system is readily driven away from the equilibrium with lattice in semiconductors. In terms of our model ($h_{e-ph}$), thermal contact between the electron- and the lattice-systems are weak in semiconductors so that they are readily driven out of equilibrium. We mention that a high mobility of electrons is often ascribed to be the cause of hot electron generation in semiconductors. The present study makes this assumption questionable; namely, a high mobility implies a high electrical conductivity (and a large $P$), but the electrical conductivity is usually higher in metals and does not explain why semiconductor is more feasible for hot electron generation.

In this study simulation calculation assumed linear transport. Namely the electrical conductivity and the thermal conductivities are assumed to be constants. In metals, nonlinear effects may not be significant since the $T_{e}$ rise is not too large. In the doped n-GaAs sample at room temperature (as in this work), nonlinear effects may not be serious up to $E\approx10kV$ \cite{Weng.Q}, so that the findings in the present study are supposed to be valid. In the higher $E$ region above $10kV/cm$, however, the electron mobility will be reduced due to the transfer of electrons to upper (X and/or L) valleys. Even in such a higher-$E$ region, our model will provide a useful guideline at the starting point.

\section{Summary}
We demonstrate that in layered conductors nonequilibrium nature between the electrons and the lattice can be explicitly pursued by separating the electron- and the lattice-subsystems into two physical layers that exchange heat at the interface. Highly nonequilibrium distribution of electrons from that of the lattice is found in a doped n-GaAs sample. In a NiCr sample with a similar configuration, the electron- and the lattice-systems are in quasi-equilibrium. Remarkable difference of the electro-thermal properties of semiconductors and metals is suggested to arise from the difference in the electron specific heat. This work provides a simple and convenient method for modeling layered conductors in a nonequilibrium condition, and will give useful hints for deeper understanding of the nonequilibrium properties of electrical conductors

\acknowledgments
We acknowledge funding support from National Key Research Program of China under grant No. 2016YFA0302000, National Natural Science Foundation of China under grant Nos. 11674070/11427807/11634012, and Shanghai Science and Technology Committee under grant Nos.18JC1420402, 18JC1410300, 16JC1400400. S.K. acknowledges support by the Chinese Academy of Sciences Visiting Professorships for Senior International Scientists.


\begin{thebibliography}{0}

\bibitem{Aninkevicius.V}
  \Name{Aninkevi\v{c}ius V., Bareikis V., Katilius R., Liberis J., Matulionien\.{e} I., Matulionis A., Sakalas P. \and \v{S}altis R.}
  \REVIEW{Phys. Rev. B}{53}{1996}{6893}.

\bibitem{Matulionis.A}
  \Name{Matulionis A., Liberis J., Matulioniene I. \and Ramonas M.}
  \REVIEW{Acta Phys. Pol., A}{113}{2008}{967}.

\bibitem{Levi.A}
  \Name{Levi A. F. J., Hayes J. R., Platzman P. M., Liberis J. \and Wiegmann W.}
  \REVIEW{Phys. Rev. Lett.}{55}{1985}{2071}.

\bibitem{Atar.F}
  \Name{Atar F. B., Aygun L. E., Daglar B., Bayindir M. \and Okyay A. K.}
  \REVIEW{Opt. Express}{21}{2013}{7196}.

\bibitem{Pumarol.M}
  \Name{Pumarol M. E., Rosamond M. C., Tovee P., Petty M. C., Zeze D. A., Falko V. \and Kolosov O. V.}
  \REVIEW{Nano Lett.}{12}{2012}{2906}.

\bibitem{Lan.Y}
  \Name{Lan Y.-W., Torres J. C. M., Zhu X., Qasem H., Adleman J. R., Lerner M. B., Tsai S.-H., Shi Y., Li L.-J., Yeh W.-K. \and Wang K. L.}
  \REVIEW{Sci. Rep.}{6}{2016}{32503}.

\bibitem{Guo.Z}
  \Name{Guo Z., Wan Y., Yang M., Snaider J., Zhu K. \and Huang L.}
  \REVIEW{Science}{356}{2017}{59}.

\bibitem{Lee.S}
  \Name{Lee S., Wijesinghe N., Diaz-Pinto C. \and Peng H.}
  \REVIEW{Phys. Rev. B}{82}{2010}{045411}.

\bibitem{Zhang.P}
  \Name{Zhang P., Fujitsuka M. \and Majima T.}
  \REVIEW{Nanoscale}{9}{2017}{1520}.

\bibitem{Li.Z}
  \Name{Li Z., Ezhilarasu G., Chatzakis I., Dhall R., Chen C.-C. \and Cronin S. B.}
  \REVIEW{Nano Lett.}{15}{2015}{3977}.

\bibitem{Lee.H}
  \Name{Lee H., Lee H. \and Park J. Y.}
  \REVIEW{Nano Lett.}{19}{2019}{891}.

\bibitem{Lee.Y}
  \Name{Lee Y. K., Jung C. H., Park J., Seo H., Somorjai G. A. \and Park J. Y.}
  \REVIEW{Nano Lett.}{11}{2011}{4251}.

\bibitem{Shao.W}
  \Name{Shao W., Yang Q., Zhang C., Wu S. \and Li X.}
  \REVIEW{Nanoscale}{11}{2019}{1396}.

\bibitem{Konstantinova.T}
  \Name{Konstantinova T., Rameau J. D., Reid A. H., Abdurazakov O., Wu L., Li R., Shen X., Gu G., Huang Y., Rettig L., Avigo I., Ligges M., Freericks J. K., Kemper A. F., D\"urr H. A., Bovensiepen U., Johnson P. D., Wang X. \and Zhu Y.}
  \REVIEW{Sci. Adv.}{4}{2018}{eaap7427}.

\bibitem{Wu.K}
  \Name{Wu K., Chen J., McBride J. R. \and Lian T.}
  \REVIEW{Science}{349}{2015}{632}.

\bibitem{Tisdale.W}
  \Name{Tisdale W. A., Williams K. J., Timp B. A., Norris D. J., Aydil E. S. \and Zhu X.-Y.}
  \REVIEW{Science}{328}{2010}{1543}.

\bibitem{Ashalley.E}
  \Name{Ashalley E., Gryczynski K., Wang Z., Salamo G. \and Neogi A.}
  \REVIEW{Nanoscale}{11}{2019}{3827}.

\bibitem{Dubi.Y}
  \Name{Dubi Y. \and Ventra M. Di.}
  \REVIEW{Rev. Mod. Phys.}{83}{2011}{131}.

\bibitem{Hartmann.M}
  \Name{Hartmann M., Mahler G. \and Hess O.}
  \REVIEW{Phys. Rev. Lett.}{93}{2004}{080402}.

\bibitem{Mahajan.R}
  \Name{Mahajan R., Chia-pin C. \and Chrysler G.}
  \REVIEW{Proceedings of the IEEE}{94}{2006}{1476}.

\bibitem{Brongersma.M}
  \Name{Brongersma M. L., Halas N. J. \and Nordlander P.}
  \REVIEW{Nature nanotechnol.}{10}{2015}{25}.

\bibitem{Peng.S}
  \Name{Peng S., Xing G. \and Tang Z.}
  \REVIEW{Nanoscale}{9}{2017}{15612}.

\bibitem{Reparaz.J}
  \Name{Reparaz J. S., Chavez-Angel E., Wagner M. R., Graczykowski B., Gomis-Bresco J., Alzina F. \and Torres C. M. S.}
  \REVIEW{Rev. Sci. Instrum.}{85}{2014}{034901}.

\bibitem{Xu.Y}
  \Name{Xu Y. N., Zhan D., Liu L., Suo H., Ni Z. H., Nguyen T. T., Zhao C. \and Shen Z. X.}
  \REVIEW{ACS Nano}{5}{2011}{147}.

\bibitem{Zhang.X}
  \Name{Zhang X., Sun D., Li Y., Lee G.-H., Cui X., Chenet D., You Y., Heinz T. F. \and Hone J. C.}
  \REVIEW{ACS Appl. Mater. Interfaces}{7}{2015}{25923}.

\bibitem{Nonnenmacher.M}
  \Name{Nonnenmacher M. \and Wickramasinghe H. K.}
  \REVIEW{Appl. Phys. Lett.}{61}{1992}{168}.

\bibitem{Harzheim.A}
  \Name{Harzheim A., Spiece J., Evangeli C., McCann E., Falko V., Sheng Y., Warner J. H., Briggs G. A. D., Mol J. A., Gehring P. \and Kolosov O. V.}
  \REVIEW{Nano Lett.}{18}{2018}{7719}.

\bibitem{Yalon.E}
  \Name{Yalon E., McClellan C. J., Smithe K. K. H., Rojo M. M., Xu R. L., Suryavanshi S. V., Gabourie A. J., Neumann C. M., Xiong F., Farimani A. B. \and Pop E.}
  \REVIEW{Nano Lett.}{17}{2017}{3429}.

\bibitem{Cui.L}
  \Name{Cui L., Jeong W., Hur S., Matt M., Klockner J. C., Pauly F., Nielaba P., Cuevas J. C., Meyhofer E. \and Reddy P.}
  \REVIEW{Science}{355}{2017}{1192}.

\bibitem{Shi.L}
  \Name{Shi L., Plyasunov S., Bachtold A., McEuen P. L. \and Majumdar A.}
  \REVIEW{Appl. Phys. Lett.}{77}{2000}{4295}.

\bibitem{Mecklenburg.M}
  \Name{Mecklenburg M., Hubbard W. A., White E. R., Dhall R., Cronin S. B., Aloni S. \and Regan B. C.}
  \REVIEW{Science}{347}{2015}{629}.

\bibitem{Weng.Q}
  \Name{Weng Q., Komiyama S., Yang L., An Z., Chen P., Biehs S.-A., Kajihara Y. \and Lu W.}
  \REVIEW{Science}{360}{2018}{775}.

\bibitem{Komiyama.S}
  \Name{Komiyama S.}
  \REVIEW{J. Appl. Phys.}{125}{2019}{010901}.

\bibitem{SINHA.S}
  \Name{SINHA S. \and GOODSON K. E.}
  \REVIEW{Int. J. Multiscale Comput. Engineering.}{3}{2005}{107}.

\bibitem{Vasileska.D}
  \Name{Vasileska D., Ashok A., Hartin O. \and Goodnick S.M.}
  \Book{Large-Scale Scientific Computing}
  \Editor{Lirkov, Ivan and Margenov, Svetozar and Wa{\'{s}}niewski, Jerzy.}
  \Publ{Springer Berlin Heidelberg}
  \Year{2010}
  \Page{451-458}.

\bibitem{Fan.A}
  \Name{Fan A., Tarau C., Bonner R., Palacios T. \and Kaviany M.}
  \REVIEW{ASME 2012 Summer Heat Transfer Conference}{2012}.

\bibitem{Garcia.S}
  \Name{Garc\'ia S., \'I\~niguez-de-la-Torre I, Mateos J., Gonz\'alez T. \and P\'erez S.}
  \REVIEW{Semicond. Sci. Technol.}{31}{2016}{065005}.

\bibitem{Wang.Y}
  \Name{Wang Y., Ao J, Liu S. \and Hao U.}
  \REVIEW{Appl. Sci.}{9}{2019}{75}.

\bibitem{Weng.QC}
  \Name{Weng Q., Lin K.-T., Yoshida K., Nema H., Komiyama S., Kim S., Hirakawa K. \and Kajihara Y.}
  \REVIEW{Int. J. Multiscale Comput. Engineering.}{3}{2005}{107}.

\end{thebibliography}
\end{document}